\documentclass[fleqn, usenatbib]{mnras}

\usepackage{newtxtext,newtxmath}

\usepackage[T1]{fontenc}
\usepackage{ae,aecompl}
\usepackage[]{units}

\usepackage{graphicx}	
\usepackage{amsmath}	
\usepackage{amssymb}	
\usepackage{graphicx}	
\usepackage{amsmath}	
\usepackage{amssymb}	
\usepackage{layouts}
\usepackage{multicol}        
\usepackage{bm}		
\usepackage{pdflscape}	
\usepackage{verbatim}

\newcommand{\be}{\begin{equation}}
\newcommand{\ee}{\end{equation}}
\newcommand{\mpcs}{\mathrm{M_\odot\,pc^{-2}}}

\newcommand{\msun}{M_{\sun}}

\newcommand{\dif}{\mathrm{d}}

\newcommand{\Smax}{\Sigma_{\rm max}}
\newcommand{\Scrit}{\Sigma_{\rm crit}}
\newcommand{\Seff}{\Sigma_{\ast,\,{\rm eff}}}

\defcitealias{hopkins:maximum.surface.densities}{Paper~I}
\defcitealias{grudic:2016.sfe}{Paper~II}

\title[Explaining the maximum stellar surface density]{The Maximum Stellar Surface Density Due to the Failure of Stellar Feedback}
\vspace{-0.5cm}

\author[Grudi\'{c} et al.]{
Michael Y. Grudi\'{c},$^{1}$\thanks{E-mail: mgrudich@caltech.edu}
Philip F. Hopkins$^{1}$,
Eliot Quataert$^{2}$,
and Norman Murray$^{3,4}$
\\
$^{1}$TAPIR, Mailcode 350-17, California Institute of Technology, Pasadena, CA 91125, USA\\
$^{2}$Department of Astronomy and Theoretical Astrophysics Center, University of California Berkeley, Berkeley, CA 94720\\
$^{3}$Canadian Institute for Theoretical Astrophysics, 60 St. George Street, University of Toronto, ON M5S 3H8, Canada \\
$^{4}$ Canada Research Chair in Astrophysics \\
}

\date{\ \ Submitted \today \vspace{-0.5cm}}

\pubyear{2018}
\begin{document}
\label{firstpage}
\pagerange{\pageref{firstpage}--\pageref{lastpage}}
\maketitle

\begin{abstract}
A maximum stellar surface density $\Smax\sim \unit[3 \times 10^5]{\mpcs}$ is observed across all classes of dense stellar systems (e.g.\ star clusters, galactic nuclei, etc.), spanning $\sim 8$ orders of magnitude in mass. It has been proposed that this characteristic scale is set by some dynamical feedback mechanism preventing collapse beyond a certain surface density. However, simple analytic models and detailed simulations of star formation moderated by feedback from massive stars argue that feedback becomes {\em less} efficient at higher surface densities (with the star formation efficiency increasing as $\sim \Sigma/\Scrit$). We therefore propose an alternative model wherein stellar feedback becomes ineffective at moderating star formation above some $\Scrit$, so the supply of star-forming gas is rapidly converted to stars before the system can contract to higher surface density. We show that such a model -- with $\Scrit$ taken directly from the theory -- naturally predicts the observed $\Smax$. $\Smax\sim 100\Scrit$ because the gas consumption time is longer than the global freefall time even when feedback is ineffective. Moreover the predicted $\Smax$ is robust to spatial scale and metallicity, and is preserved even if multiple episodes of star formation/gas inflow occur. In this context, the observed $\Smax$ directly tells us where feedback fails.
\end{abstract}

\begin{keywords}
galaxies: formation -- galaxies: evolution -- galaxies: active -- galaxies: star formation -- cosmology: theory -- galaxies: star clusters: general\vspace{-0.7cm}
\end{keywords}



\section{Introduction}
\label{sec:intro}

\citet{hopkins:maximum.surface.densities} (hereafter \citetalias{hopkins:maximum.surface.densities}) showed that the central surface densities of essentially all dense stellar systems exhibit the same apparent upper limit $\Smax \sim \unit[3\times10^5]{\mpcs}$. This includes globular clusters (GCs), super star clusters (SSCs), dwarf and late-type galaxy nuclear star clusters (NSCs), young massive clusters (YMCs), ultra-compact dwarfs (UCDs), compact ellipticals (cEs), galactic bulges, nearby and high-redshift early-type/elliptical galaxies, even nuclear stellar disks around Sgr A$^{\ast}$ and the Andromeda nuclear black hole. These span mass scales of $10^{4}-10^{12}\,M_{\odot}$, spatial sizes $\unit[0.1-10^4]{pc}$, {\em three-dimensional} densities $1-10^{5}\,M_{\sun}\,{\rm pc^{-3}}$ (free-fall times $\sim 10^{4}-10^{7}$\,yr), $N$-body relaxation times $\sim 10^{6}-10^{17}$\,yr, escape velocities $\sim 20-600\,{\rm km\,s^{-1}}$, metallicities $Z\sim 0.01-5\,Z_{\odot}$, and formation redshifts $z\sim 0-6$, yet agree in $\Smax$.

\begin{figure*}
\includegraphics[width=\textwidth]{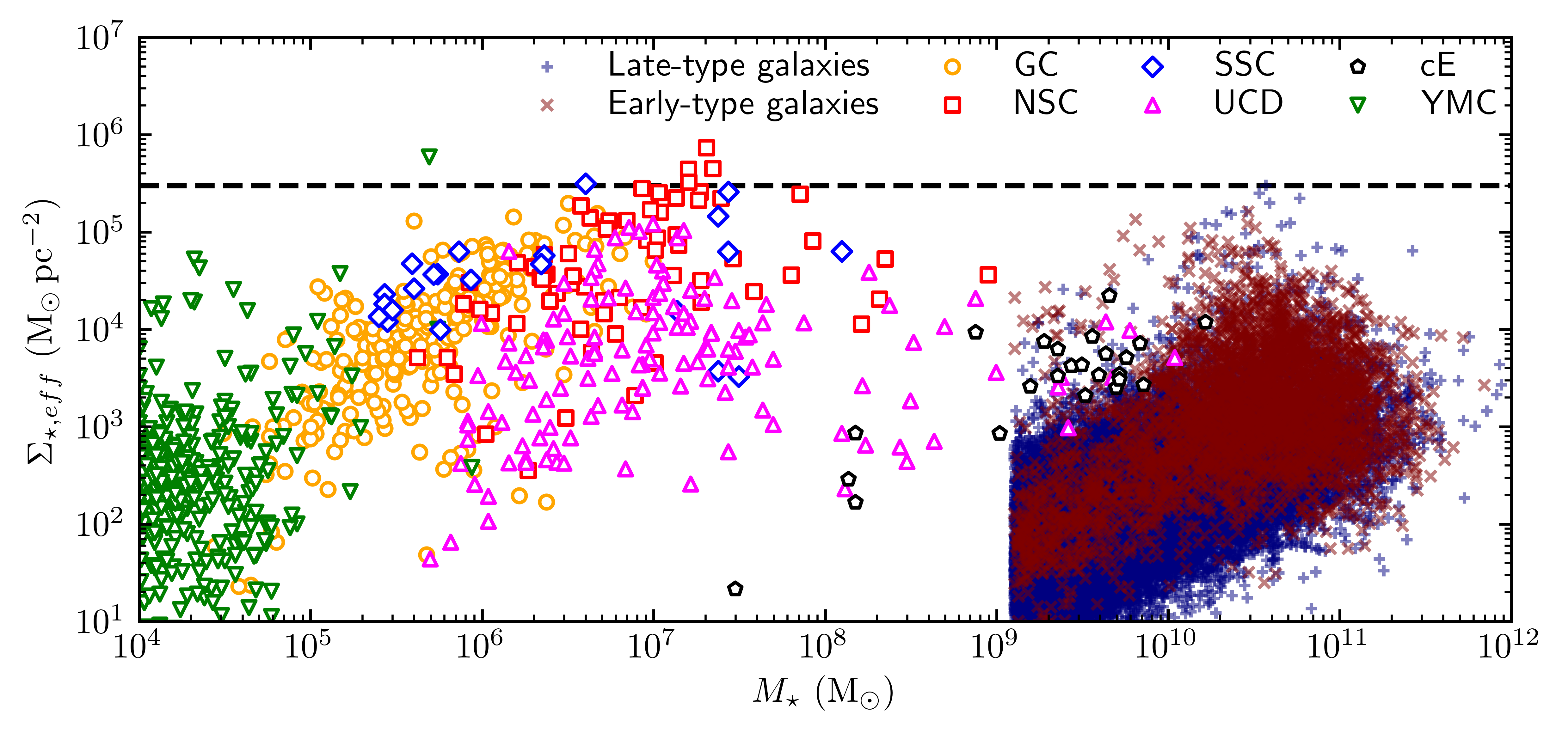}
\vspace{-0.75cm}
\caption{``Effective'' stellar surface density ($\Seff \equiv M_{\ast} / (2\pi\,R_{\rm eff}^{2})$) as a function of stellar mass for various types of stellar systems. Late- and early-type galaxies range from redshifts $z=0-3$ and are taken from \citet{vdw:2014.candels}. Globular clusters (GC), nuclear star clusters (NSC), ultra-compact dwarfs (UCD), and compact ellipticals (cE) are from the compilation of \citet{aimss}. Super star clusters (SSC) are from the populations in M82 \citep{mccrady:m82.sscs}, NGC 7252 \citep{bastian:2013.ngc7252.clusters}, NGC 34 \citep{schweizer:2007.ngc30.ssc}, and NGC 1316 \citep{bastian:2006.sscs}. Young massive clusters (YMC) are from the Milky Way \citep{pz2010} and M83 \citep{ryon:2015.m83.clusters} populations. {\it Dashed:} Fiducial maximum effective surface density $\Smax=\unit[3\times10^5]{\mpcs}$.}\vspace{-0.25cm}
\label{fig:mr}
\end{figure*}

\begin{figure}
\includegraphics[width=\columnwidth]{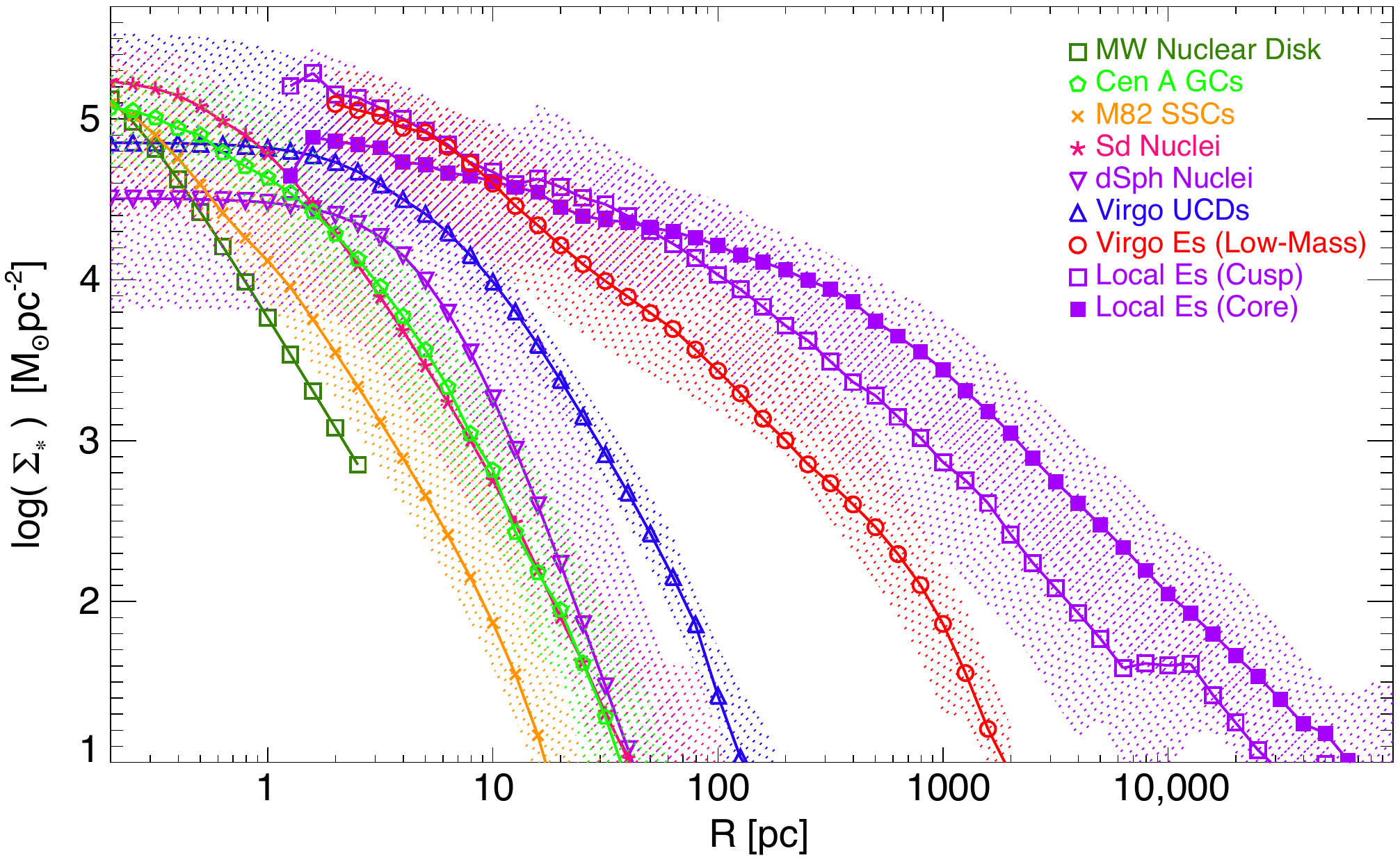}
\vspace{-0.45cm}
\caption{Observed stellar surface density {\em profiles} $\Sigma_{\ast}(r)$ as a function of projected radius, within individual stellar systems -- reproduced from Fig.~2 of \citetalias{hopkins:maximum.surface.densities}. Lines show the median $\Sigma_{\ast}(r)$ from each sample, shaded range the $\pm 1\,\sigma$ range in $\Sigma_{\ast}(r)$ from all profiles in the sample. Samples are: Milky Way nuclear stellar disk \citep{lu:mw.nuclear.disk}, Cen A GCs \citep{rejkuba:cenA.gcs}, M82 SSCs \citep{mccrady:m82.sscs}, NSCs in late-type (Sd) galaxy nuclei \citep{boker04:nuclei.scalings}, NSCs in dwarf-Spheroidal galaxy nuclei \citep{geha02:dE.nuclei}, UCDs in Virgo \citep{evstigneeva:virgo.ucds}, early-type galaxies in Virgo (separated into low-mass ``dwarf ellipticals'' from \citealt{jk:profiles}, and massive ``cusp''/steep profile or ``core''/shallow-profile systems from \citealt{lauer:bimodal.profiles}). Although many of these (e.g.\ the massive early-type galaxies) have $\Seff$ (defined at large radii $\gtrsim$\,kpc) well below $\Smax$, all systems appear to approach (and where resolved, saturate around) the fiducial maximum surface density $\Smax=\unit[3\times10^5]{\mpcs}$.}\vspace{-0.25cm}
\label{fig:profiles}
\end{figure}

In Figure \ref{fig:mr} we compile more recent observations of dense stellar systems of all classes, and find that this still holds largely true, although some nuclear star clusters exceeding the fiducial value of $\Smax$ by a factor of a few have since been found. Figure \ref{fig:profiles} is adapted from the original compilation of mass profiles of individual objects in \citetalias{hopkins:maximum.surface.densities} -- it shows that even many systems with ``effective'' surface densities (measured at $R_{\rm eff}$) have {\em central} surface densities which approach but do not appear to exceed $\Smax$ (at least where resolved).

\citetalias{hopkins:maximum.surface.densities} speculated that the universality of $\Smax$ might owe to stellar feedback processes.\footnote{They also discussed some possible explanations related to e.g.\ mergers, angular momentum transport processes, or dynamical relaxation, which they showed could {\em not} explain $\Smax$ across the wide range of systems observed (e.g. dynamical relaxation cannot dominate the systems with relaxation times much longer than a Hubble time, and global processes unique to galaxy mergers cannot explain star cluster interiors).} After all, it is widely-recognized that feedback plays an important role regulating star formation (SF) in cold, dense molecular clouds \citep[see][for a review]{kennicutt:2012.review}. As gas collapses and forms stars, those stars inject energy and momentum into the ISM via protostellar heating and outflows, photoionization and photoelectric heating from UV photons, stellar winds, radiation pressure and supernova explosions. All of these mechanisms may moderate SF, either by contributing to the disruption of molecular clouds \citep{larson:gmc.scalings,murray:molcloud.disrupt.by.rad.pressure, hopkins:fb.ism.prop, krumholz:2014.feedback.review, grudic:2016.sfe} or the large-scale support of galaxies against vertical collapse \citep{thompson:rad.pressure,ostriker.shetty:2011, cafg:sf.fb.reg.kslaw,hopkins:2013.fire,orr:2016.what.fires.up.SF}. These mechanisms have various roles on different scales, but stellar feedback is generally is the only force strong enough to oppose gravity in dense, star-forming regions, so the characteristic scale of a newly-formed stellar system should be determined by the balance point of feedback and gravity. 

The specific possibility discussed in \citetalias{hopkins:maximum.surface.densities} was that multiple-scattering of IR photons might build up radiation pressure to exceed the Eddington limit for dusty gas. However, the value of $\Smax$ predicted according to this argument is inversely proportional to metallicity, so does not explain why $\Smax$ is apparently the same in SSCs in metal-rich starbursts \citep{keto:2005.m82.gmcs,mccrady:m82.sscs} (or super-solar massive elliptical centers) and in metal-poor GCs (or metal-poor high-$z$, low-mass compact galaxies). The argument therein also relied on scalings between IR luminosity and star formation rate (SFR) valid only for continuous-star forming populations with duration longer than $\sim 10-30\,$Myr, which exceeds the dynamical times of many of these systems. Finally, \citet{aimss} noted that this effect cannot prevent the system from exceeding $\Smax$ if SF occurs in multiple episodes.

Since then, various theoretical works have noted the importance of surface density in setting the ratio between the momentum-injection rate from massive stars and the force of self-gravity in a star forming cloud \citep{fall:2010.sf.eff.vs.surfacedensity,murray:molcloud.disrupt.by.rad.pressure,dekel:2013.giant.clumps, thompson:2016.eddington.outflows,raskutti:2016.gmcs,grudic:2016.sfe}. For a cloud with total mass $M$ and stellar mass $M_\star=\epsilon_{\rm int} M $,
\begin{equation}
\label{eqn:fgrav.fsf}\frac{F_{\rm gravity}}{F_{\rm feedback}} \sim \frac{\frac{GM^2}{R^2}}{\epsilon_{\rm int} M \langle \frac{\dot{P}_\star}{M_\star} \rangle} \sim \frac{\Sigma}{\Scrit},
\end{equation}
where $\langle \frac{\dot{P}_\star}{M_\star} \rangle$ is the specific momentum injection rate from stellar feedback assuming a simple stellar population with a well-sampled IMF, which is $\sim \unit[10^3]{\frac{L_\odot}{M_\odot c}}$ for the first $\unit[3]{Myr}$ after SF, and $\Scrit \sim \langle\frac{\dot{P}_\star}{M_\star}\rangle /G \approx 3000\,M_{\sun}\,{\rm pc^{-2}}$ is the characteristic surface density that parametrizes the strength of feedback. If the final SF efficiency (SFE) $\epsilon_{\rm int}$ is ultimately set by the balance of feedback and gravity, one expects that $\epsilon_{\rm int}\rightarrow 1$ for $\Sigma \gg \Scrit$ \citep{fall:2010.sf.eff.vs.surfacedensity}. The detailed simulations of \citet{grudic:2016.sfe} \citepalias{grudic:2016.sfe} showed that this argument is valid across a wide range of metallicities, surface densities and spatial scales, and the final SFE of a molecular cloud is a function mainly of $\Sigma$, with weak dependence upon other factors. \citetalias{grudic:2016.sfe} also found that the final ratio of stellar mass to initial gas mass, $\epsilon_{\rm int}$, is proportional to the fraction of gas converted to stars within a freefall time, $\epsilon_{\rm ff}$, because a GMC always tends to form enough stars to destroy itself within a few freefall times. Thus, $\Sigma$ should parametrize the per-freefall efficiency of SF in a manner insensitive to spatial scale and metallicity.

In this paper, we show that if gas contracts globally (for any reason), as it becomes denser ($\Sigma$ increases), and gravity becomes stronger relative to stellar feedback, gas is converted more and more rapidly into stars (above a characteristic surface density $\Scrit$). This exhausts the gas supply, preventing any significant fraction of the inflow from reaching surface densities $>\Smax$. We calculate $\Smax$ in terms of $\Scrit$ and show that the observed $\Smax\sim \unit[3\times 10^5]{\mpcs}$ is naturally predicted by the value $\Scrit = \unit[3000]{\mpcs}$ set by feedback from massive stars \citep{fall:2010.sf.eff.vs.surfacedensity, grudic:2016.sfe}.

\vspace{-0.5cm}
\section{Derivation}
\label{sec:deriv}
\begin{figure}
\includegraphics[width=\columnwidth]{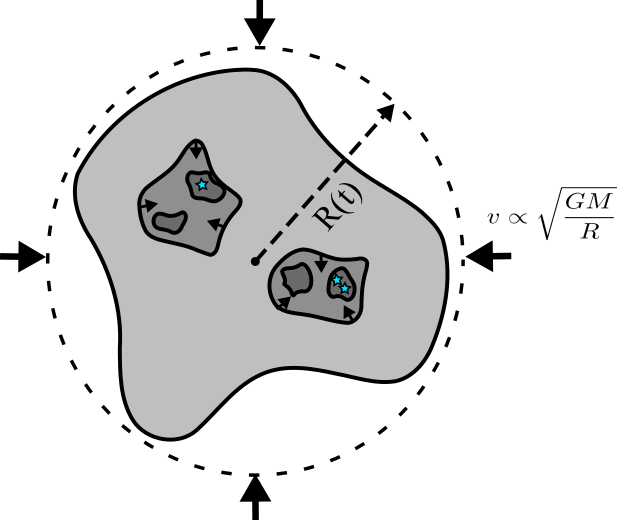}
\vspace{-0.25cm}\caption{Schematic of our proposed ``best-case'' scenario for the formation of a dense stellar system (\S~\ref{sec:deriv}). A star-forming gas cloud of initial gas mass $M$ is localized within a sphere of radius $R$. It collapses coherently at the freefall velocity $v_{\rm ff}=\sqrt{\frac{2GM}{R}}$, while fragmenting locally and forming stars in dense sub-regions. In this ``best case'', no dynamical mechanism slows the collapse significantly.}\vspace{-0.25cm}
\label{fig:cartoon}
\end{figure}

Consider a discrete SF episode involving a finite collapsing gas mass $M$, as illustrated in Figure \ref{fig:cartoon}. At a give time, the mass is localized within a radius $R$, so that its {\em mean} surface density is:
\begin{equation}
\Sigma=\frac{M}{\pi R^2}.
\end{equation}
It is forming stars at some SFR, which we can parameterize with the SFE:
\begin{equation}
\text{SFR} \equiv \frac{\dif M_{\ast}}{\dif t} = \frac{\epsilon_{\rm ff} M_{\rm gas} }{t_{\rm ff}},
\end{equation}
where $\epsilon_{\rm ff}$ is the (possibly variable) per-freefall SFE and $t_{\rm ff} = \frac{\pi}{2} \sqrt{\frac{R^3}{2 G M}}$ is the freefall time. 

Now, since we are only interested in the {\em maximum} stellar surface density such a system might reach, we will assume the ``best-case'' scenario for forming a dense stellar system. Specifically, assume:
\begin{enumerate}
\item The gas cloud is collapsing at a speed on the order of the escape velocity:
\begin{equation}
\frac{\dif R}{\dif t}  = - x_{ff}\sqrt{\frac{2GM}{R}},
\end{equation}
where $x_{ff}$ is a constant of order unity.
\item There is no support against collapse from large-scale turbulent motions,\footnote{Note that some amount of turbulence must be assumed if stars are forming. We assume that such turbulent eddies are small compared to $R$, and thus are advected with the large-scale collapse without strongly opposing it.} tidal forces, rotation, shear, magnetic fields, cosmic rays, or the dynamical effects of stellar feedback. We neglect all of these because we are interested in the best-case scenario for producing a dense stellar system according to a give SFE law -- any of these may be present, but they will only {\em slow} collapse, making a lower-density system in the end.
\end{enumerate}
This is an idealization, but \citet{kim:2017.globular.clusters} did find that bound star clusters do form in a coherent collapse with velocity on the order of the freefall velocity in cosmological simulations, and stellar feedback does not greatly affect the dynamics until a significant fraction of the gas mass has been converted to stars.

We shall assume that $\epsilon_{\rm ff}$ has some explicit dependence upon $\Sigma$, as is motivated by previous work. Accounting for radiation pressure, stellar winds, photoionization heating, and SN explosions, \citetalias{grudic:2016.sfe} found
\begin{equation}
\epsilon_{\rm ff} = \epsilon_{\rm ff}(\Sigma) = \left(\frac{1}{\epsilon_{\rm ff}^{\rm max}} + \frac{\Scrit}{\Sigma} \right)^{-1},
\label{eq:eff}
\end{equation}
where $\Scrit=\unit[3000]{\mpcs}$ is set by the strength of these feedback mechanisms. The dimensionless quantity $\epsilon_{\rm ff}^{\rm max}$ is the maximum per-freefall SFE attained as $\Sigma \rightarrow \infty$. In star-forming clouds supported at a fixed mean surface density, \citetalias{grudic:2016.sfe} (see their Eq.~13 and Fig.~5) found that $\epsilon_{\rm ff} \approx 0.34\,\epsilon_{\rm int}$ (where $\epsilon_{\rm int}$ is the fraction of gas turned into stars over the entire integrated SF history, which of course just saturates at $\epsilon_{\rm int}^{\rm max}=1$ as $\Sigma \rightarrow \infty$). However, this was the median over the entire SF history including initial collapse and eventual blowout. Therefore  -- in our ``best-case'' coherent collapse scenario, we are only interested in the ``peak SFR'' event, so $\epsilon_{\rm ff}^{\rm max}$ should be somewhat greater, $\sim 0.5$ (see \citetalias{grudic:2016.sfe}, Figure 3), and subject to further order-unity corrections due to the different collapse geometry from these simulations. In general, $\epsilon_{\rm ff}^{\rm max}$ should be similar that predicted by turbulent molecular cloud simulations that do not include stellar feedback, which have generally found $\epsilon_{ff} \sim 0.5-1$ in the limit of large turbulent Mach number and realistic turbulent forcing \citep{federrath:2012.sfr.vs.model.turb.boxes}.


The SFR of the cloud is
\begin{equation}
\begin{split}
\frac{\dif M_{\ast}}{\dif t}  &= \epsilon_{\rm ff}\frac{ M_{\rm gas}}{\,t_{\rm ff}} = \frac{M_{\rm gas}}{\left(\epsilon_{\rm ff}^{\rm max} \right)^{-1} + \frac{\Scrit}{\Sigma}} \sqrt{\frac{8 G M}{\pi^2 R^3}}, \\
\end{split}
\end{equation}
where $M_{\rm gas}$ is the gas mass that has not been converted to stars at time $t$. The differential equation for the gas mass converted to stars when the cloud has radius $R$ follows:
\begin{equation}
\begin{split}
\frac{\dif M_{\rm gas}}{\dif R} &= -\frac{\dif M_{\ast}}{\dif t} \, \frac{\dif t}{\dif R} = -\frac{2}{\pi x_{ff}R}\frac{M_{\rm gas}}{\left(\epsilon_{\rm ff}^{\rm max} \right)^{-1} + \frac{\pi R^2\Scrit}{M}}.
\end{split}
\label{eq:diffeq}
\end{equation}
The solution for the fraction of the gas mass surviving to radii $<R$ is
\begin{equation}
\frac{M_{\rm gas}\left(<R\right)}{M} = \left(1 - \frac{M}{M+\epsilon_{\rm ff}^{\rm max} \pi R^2 \Sigma_{crit}}\right)^{\epsilon_{\rm ff}^{\rm max}/\pi x_{ff}}.
\label{eq:mr}
\end{equation}
\begin{figure}
\includegraphics[width=\columnwidth]{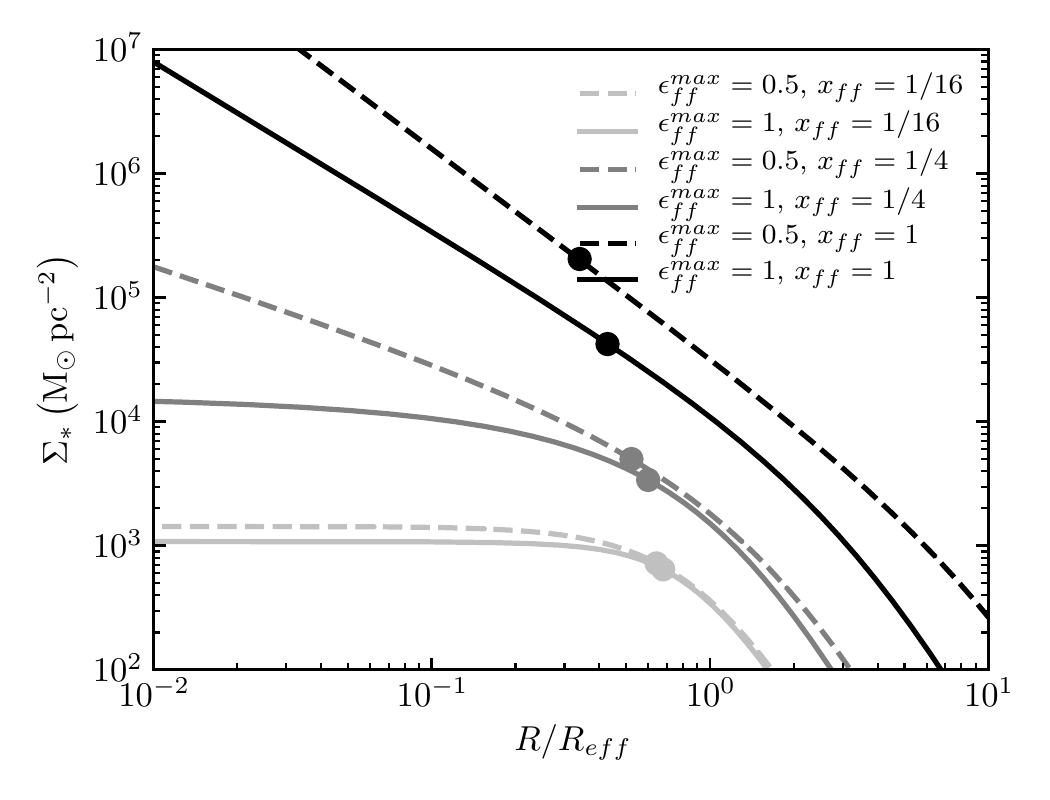}
\vspace{-0.6cm}
\caption{Radial stellar surface density profiles computed from the cloud collapse model described in \S~\ref{sec:deriv} for various values of the maximum per-freefall SFE $\epsilon_{\rm ff}^{\rm max}$ and the rate of collapse relative to freefall $x_{ff}$, with radius in units of the half-mass radius $R_{eff}$. The point on each curve gives the effective stellar surface density $\Sigma_{\ast,eff}=M_\ast/\left(2\pi R_{eff}^2\right)$ of the model. The characteristic surface densities obtained over the parameter ranges $\epsilon_{\rm ff}^{\rm max} \sim 0.5-1$ and $x_{ff} \sim 0.1-1$ span the range $10^3-10^6 \,M_{\sun}\,{\rm pc^{-2}}$ in which most dense stellar systems lie (see Fig. \ref{fig:mr}). To form a system with $\Sigma_{\ast,eff}>>3\times 10^5 \,M_{\sun}\,{\rm pc^{-2}}$ would require $\epsilon_{\rm ff}^{\rm max} << 0.5$ or $x_{ff} >>1$, both of which are unphysical.}
\label{fig:modelprofile}
\end{figure}
Thus, as $R\rightarrow 0$ we see that $M_{\rm gas} \rightarrow 0$, ie. the gas is exhausted as the system contracts to surface densities $\Sigma \gg\Scrit$. The stellar system formed will subsequently undergo a period of relaxation, but energy conservation requires that the stars remain on orbits with apocentres on the order of the radius $R$ at which they formed \footnote{We have verified with collisionless Monte Carlo simulations that the functional form Equation \ref{eq:mr} does closely match the final stellar mass distribution after violent relaxation to virial equilibrium, provided that the initial virial parameter $2 E_{kin}/|E_{grav}| \sim 1$.}. We may thus construct a radial stellar density profile as the superposition of the top-hat mass distributions formed at each radius. The corresponding projected stellar surface density profile is
\begin{equation}
\Sigma_\ast \left(R\right) = 2 \int_R^\infty \sqrt{R'^2 - R^2} \frac{\dif M_{gas}\left(<R'\right)}{\dif R'}/\left(\frac{4 \pi}{3}R'^3\right) \dif R',
\end{equation}
which we plot for various values of $\epsilon_{\rm ff}^{\rm max}$ and $x_{ff}$ in Figure \ref{fig:modelprofile}. In general, we find that the characteristic stellar surface densities for plausible values of $\epsilon_{\rm ff}^{\rm max}$ and $x_{\rm ff}$ span the range of surface densities found in dense stellar systems (Figs. \ref{fig:mr} and \ref{fig:profiles}). Furthermore, if $x_{\rm ff}=1$ then effective surface densities $\sim 10^5 \rm{M_\odot\,pc^{-2}}$ are obtained, corresponding to the maximum observed.

It should be noted that the inner surface density profile plotting in Figure \ref{fig:modelprofile} is $\Sigma \propto R^{-2 + \frac{2 \epsilon_{\rm ff}^{\rm max}}{\pi x_{\rm ff}}}$, which is nearly as steep as $R^{-2}$ for the physically-plausible parameters $\epsilon_{\rm ff}^{\rm max} = 0.5$ and $x_{\rm ff}=1$, ie. the profile has nearly constant mass per interval in $\log R$. In such a case a non-negligible fraction of the mass can be concentrated on scales $<0.1 {\rm pc}$. Such a high degree of central concentration is not generally found in any type of stellar system, so the inner profiles in this model are an unphysical artifact of the imposed condition of unopposed, spherically-symmetric collapse. This is never realized in nature because even an initially-monolithic supersonic collapse is unstable to fragmentation (Guszejnov et al. 2018, in prep.), and the subsequent violent relaxation of stars produces a much shallower (typically flat) inner density profile \citep{klessen.burkert:2001, bonnell:2003.hierarchical, grudic:cluster.properties}. Thus, our free-collapse model lacks the physics necessary to establish a hard limit upon the central stellar surface density \footnote{Indeed, there is at least one YMC in M83 with central surface density in excess of $10^6 \msun\,{\rm pc^{-2}}$ in the catalogue of \citep{ryon:2015.m83.clusters}, suggesting that the same bound for {\it central} surface density might not strictly hold.}, although it should scale in a similar way to the effective surface density when combined with the action of the scale-free physics of gravity and turbulence during star formation.

\begin{figure}
\centering
\includegraphics[width=\columnwidth]{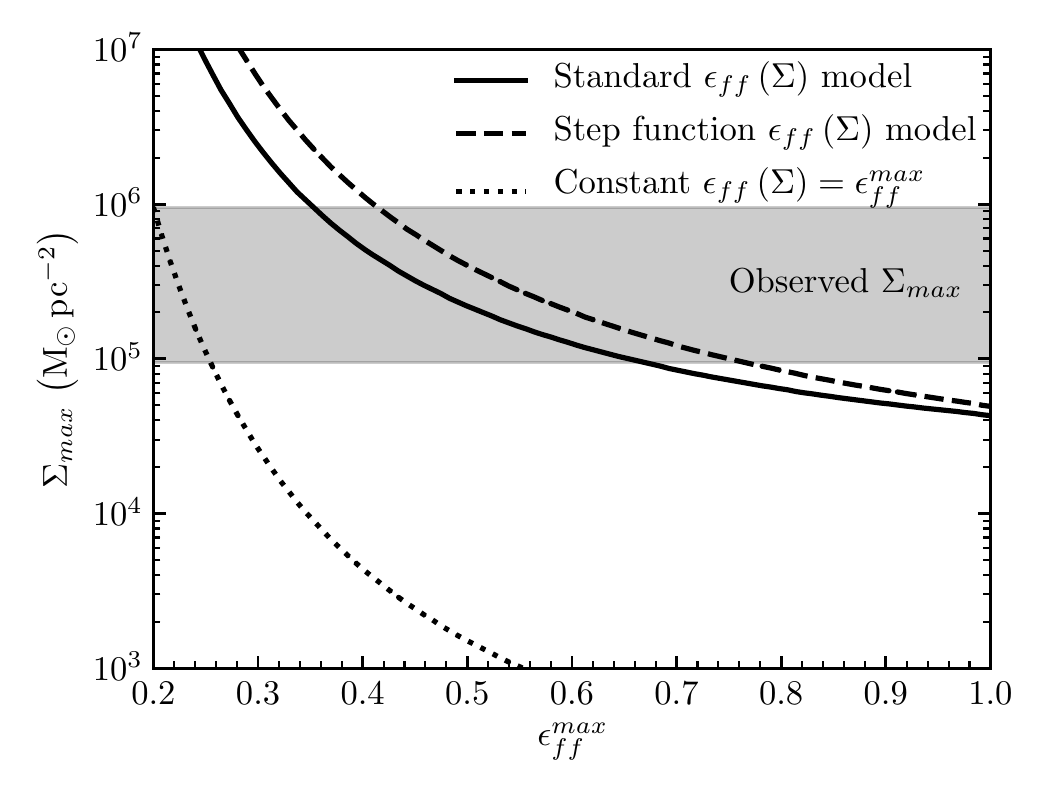}
\vspace{-0.6cm}
\caption{Maximum effective surface density $\Smax$ predicted by the model in \S~\ref{sec:deriv} as a function of the parameter $\epsilon_{\rm ff}^{\rm max}$ (maximum SFE per free-fall time, as $\Sigma\rightarrow \infty$), assuming $\Scrit\approx3000\,\msun\,{\rm pc^{-2}}$. Simulations give $\epsilon_{\rm ff}^{\rm max} \approx 0.5-1$. 
Different lines compare different models for how the efficiency $\epsilon_{\rm ff}$ scales at finite $\Sigma$. 
{\em Solid}: Our fiducial model (Eq.~\ref{eq:eff}), where $\epsilon_{\rm ff}$ scales with $\Sigma/\Scrit$ as expected from simple analytic comparison of feedback and gravity (Eq.~\ref{eqn:fgrav.fsf}) or detailed SF simulations (\citetalias{grudic:2016.sfe}). 
{\em Dashed}: A model where $\epsilon_{\rm ff}$ scales as a step function, with  $\epsilon_{\rm ff}=0.01$ when $\Sigma<\Scrit$, and $\epsilon=\epsilon_{\rm max}$ when $\Sigma < \Scrit$. This gives similar results to the fiducial case, demonstrating that the details of {\em how} $\epsilon_{\rm ff}$ scales do not matter here, so long as it rises efficiently above $\sim \Scrit$. 
{\em Dotted}: A model with constant $\epsilon_{\rm ff} = \epsilon_{\rm ff}^{\rm max}$, {\em independent} of surface density $\Sigma$. This gives a very steep dependence and can only be reconciled with the observed $\Smax$ if we fine-tune $\epsilon_{\rm ff}^{\rm max}$ to a value outside the range predicted by numerical simulations.}
\vspace{-0.25cm}
\label{fig:sigmamax}
\end{figure}

In Figure \ref{fig:sigmamax}, we consdier the maximally-freefalling case $x_{\rm ff}=1$ to plot the depedence of  $\Smax$ on $\epsilon_{\rm ff}^{\rm max}$. We find that if $\Scrit=\unit[3000]{\mpcs}$ and the plausible range for $\epsilon_{\rm ff}^{\rm max}$ is $0.5-1$, the predicted $\Smax$ lies within an order of magnitude of the observed $\Smax\sim\unit[3\times 10^5]{\mpcs}$ (Figure \ref{fig:mr}). We also present results for two alternate models for $\epsilon_{\rm ff}(\Sigma)$: a constant value, and a step-function equal to $0.01$ \citep[e.g.][]{kennicutt98,krumholz:2012.universal.sf.efficiency} below $\Scrit$ and $\epsilon_{\rm ff}^{\rm max}$ above $\Scrit$. First, we note that while our preferred model gives $\Sigma_{\rm max}$ independently of initial cloud surface density, these do not -- we therefore take the initial density to be $\unit[100]{\mpcs}$, typical of local GMCs \citep{bolatto:2008.gmc.properties}. Second, we see the ``$\epsilon_{\rm ff}=$\,constant'' model predicts a $\Smax$ that is more sensitive to the chosen $\epsilon_{\rm ff}$ (and the ``preferred'' value, $\sim 0.2$, is small). The step-function model, however, gives very similar results to our default model, so we see that the conclusions are not specific to the {\em details} of how $\epsilon_{\rm ff}$ scales, so long as $\epsilon_{\rm ff}$ is small when $\Sigma<\Scrit$ and grows to a value of order unity above $\Sigma \sim \Scrit$. Ultimately, the $\unit[2]{dex}$ separation between $\Scrit$ and $\Smax$ can be understood as follows: the system forms stars slowly until reaching $\Sigma \sim \Sigma_{crit}$, and only then does significant star formation happen, during which global collapse still proceeds. Thus this system is significantly denser than $\Sigma_{crit}$ at the median star formation time. 

\vspace{-0.6cm}
\section{Discussion}

%

We have shown that the observed, apparently universal maximum stellar surface density of dense stellar systems is a natural consequence of feedback-regulated SF physics. Specifically, assuming standard stellar evolution and feedback physics (from e.g.\ the combination of stellar winds, radiation pressure, SNe, etc.), then as surface densities ($\Sigma$) increase, the strength of gravity relative to feedback (assuming some fixed fraction of gas has turned into stars) increases in direct proportion to $\Sigma$ (Eq.~\ref{eqn:fgrav.fsf}; see references in \S~\ref{sec:intro}). Essentially, the strength of gravity scales $\propto G\,M^{2}/R^{2} \propto M\,\Sigma$, while the strength of feedback is proportional to the number of massive stars $\propto M$. So SF becomes more efficient, until the gas depletion timescale becomes comparable to the free-fall time, and the gas is exhausted before it can collapse to yet higher densities (even if it is getting denser as rapidly as possible, by collapsing at the escape velocity). Adopting standard scalings for the efficiency of feedback from simulations of star-forming clouds that span the relevant range of densities (\citetalias{grudic:2016.sfe}), we show this predicts a $\Smax$ in good agreement with that observed. 

This explanation has several advantages over the previously-proposed explanations of the maximum surface density from \citetalias{hopkins:maximum.surface.densities}. As \citetalias{grudic:2016.sfe} found that the parameters $\epsilon_{\rm ff}^{\rm max}$ and $\Scrit$ were insensitive to spatial scale below $\sim \unit[1]{kpc}$, our explanation applies equally well across the {\em entire} range of sizes of observed stellar systems in Figure \ref{fig:mr}. \citetalias{grudic:2016.sfe} also found SFE to be relatively insensitive to metallicity, so the $\Smax$ we calculate is not specific to a particular metallicity. The main effect of metallicity in is the aforementioned opacity to reprocessed FIR emission, but radiation hydrodynamics simulations of SF in the IR-thick limit \citep{skinner:2015.ir.molcloud.disrupt,tsang:2017.ssc.rp} have shown that this can only reduce $\epsilon_{\rm ff}$ by $\sim 30\%$, down to levels consistent with \citetalias{grudic:2016.sfe}. At fixed $\Sigma$, this explanation is also insensitive to the three-dimensional density, $N$-body relaxation time, formation redshift, and escape velocity of the stellar systems (see e.g.\ Fig.~4 in \citetalias{grudic:2016.sfe}).

This model also explains why SF in a pre-existing dense stellar system does not generally drive $\Sigma_{\ast}$ beyond $\Smax$ -- in other words, if one continuously or repeatedly ``trickled'' gas into e.g.\ a galaxy center, why could one not continuously add new stars to the central cusp, eventually exceeding $\Smax$? The key here is that the pre-existing stellar mass still contributes to the binding force of gravity: recall, $\Sigma$ in our model is the {\em total} mass, of gas+stars. This drives up the SFE whenever the {\em total} surface density exceeds $\Scrit$. Thus, for example, if fresh gas falls coherently into the centre of a bulge or dwarf nucleus with $\Sigma \sim \Smax$, then the total surface density will exceed $\Scrit$ out at larger radii, driving the SFE to high values and exhausting the gas. Multiple SF episodes would therefore be expected to build up the stellar mass by {\it increasing the radius inside of which $\Sigma \sim \Smax$}, {\em not} by increasing $\Smax$.

We also stress, of course, that $\Smax$ is not a ``hard'' limit, either in observations (Figs.~1-2), or in our model (Fig.~4). {\em Some} gas can survive to reach higher densities (and must, to fuel super-massive black holes, for example), and some gas may be re-injected by stellar mass loss in these dense nuclei. And the key parameters of our model (the efficiency of feedback, which appears in $\Scrit$, and $\epsilon_{\rm ff}^{\rm max}$) are not expected to be {\em precisely} universal, as e.g.\ variations in IMF sampling (since massive stars dominate the feedback) will alter $\Scrit$ and the exact geometry of collapse will alter $\epsilon_{\rm ff}^{\rm max}$ (at the tens of percent level). 

\section*{Acknowledgements}
We thank Arjen van der Wel for providing the galaxy size and mass data from \citet{vdw:2014.candels}. Support for MG and PFH was provided by an Alfred P. Sloan Research Fellowship, NASA ATP Grant NNX14AH35G, and NSF Collaborative Research Grant \#1411920 and CAREER grant \#1455342. Numerical calculations were run on the Caltech compute clusters `Zwicky' (NSF MRI award $\#$ PHY-0960291) and `Wheeler'. 


\vspace{-0.6cm}
\bibliographystyle{mnras}
\bibliography{master} 





\bsp	
\label{lastpage}
\end{document}